\documentclass{ws-ijmpd}
\usepackage[super,compress]{cite}
\begin{document}

\markboth{Zaza Osmanov} {}

%
\catchline{}{}{}{}{}
%

\title{On the role of the curvature drift instability in the
dynamics of electrons in active galactic nuclei
}

\author{ZAZA OSMANOV}

\maketitle

\begin{history}
\received{Day Month Year}
\revised{Day Month Year}
\end{history}

\begin{abstract}
We study the influence of the centrifugally driven curvature drift
instability (CDI) on the dynamics of relativistic electrons in the
magnetospheres of active galactic nuclei (AGN). We generalize our
previous work by considering relativistic particles with different
{\it initial phases}. Considering the Euler, continuity, and
induction equations, by taking into account the resonant conditions,
we derive the growth rate of the CDI. We show that due to the
centrifugal effects, the rotational energy is efficiently pumped
directly into the drift modes, that leads to the generation of a
toroidal component of the magnetic field. As a result, the magnetic
field lines transform into such a configuration when particles do
not experience any forces and since the instability is centrifugally
driven, at this stage the CDI is suspended.
\end{abstract}

\keywords{Magnetohydrodynamics; Plasmas; Galactic; Nuclei;
Acceleration}

\ccode{PACS numbers: 95.30.Qd; 98.62.Js; 94.20.wc}


\section{Introduction}
Usually in AGN magnetospheres the magnetic field varies in the
following interval $10^2-10^4$G depending on the distance from the
black hole event horizon. Therefore, magnetic field is very strong
and leads to the frozen-in condition of plasmas. This means that
plasma particles are forced to follow magnetic field lines. On the
other hand, AGN magnetospheres are characterized by rotational
motion and hence, it is clear that one has to study how the plasma
goes through the light cylinder surface (LCS) (a hypothetical zone
where the linear velocity of rigid rotation exactly equals the speed
of light). It is evident that if the rigid rotation is preserved,
sooner or later the physical system will encounter violation of the
causality principle. Therefore, in the mentioned zone, a certain
twisting process of magnetic field lines must exist, by means of
which, the particles will lag behind the rotation avoiding the
aforementioned problem. In Ref. \refcite{r03} authors considered
curved trajectories and generalized a work developed by Machabeli \&
Rogava (see Ref. \refcite {mr}). The authors examined a single
particle, sliding along a corotating, curved channel. In this simple
mechanical model the channel plays a role of magnetic field lines.
It was shown that dynamics of particles asymptotically becomes
force-free if a shape of the channel is given by the Archimedes'
spiral and the particles cross the LCS without violating the
causality principle.

\begin{figure}
\par\noindent
{\begin{minipage}[t]{1.\linewidth}
\includegraphics[width=\textwidth] {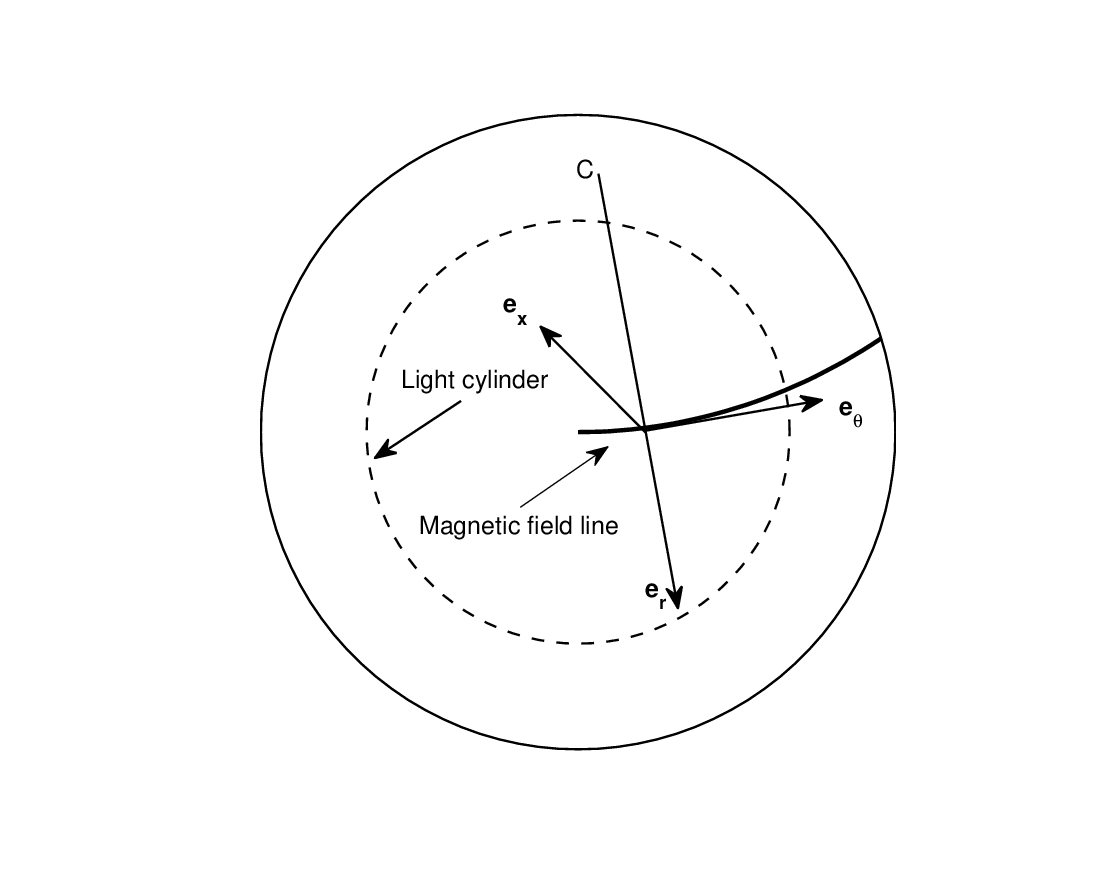}
\end{minipage}
 }
 \caption[ ] {The geometry in
which the set of main Eqs. (\ref{eul}-\ref{ind}) is considered;
(${\bf e}_{\theta},{\bf e}_r,{\bf e}_x$) denotes orthonormal basis
of unit vectors; ${\bf e}_x$ is directed perpendicularly to the
plane of the figure (equatorial plane); $C$ is the center of the
curvature; and ${\bf e}_{\theta}$ and ${\bf e}_r$ are the tangential
and perpendicular (with respect to the field line) unit vectors
respectively.}\label{fig1a}
 \end{figure}

Generally speaking, if the magnetic field is still robust, one has
to twist the field lines in an appropriate way. Therefore, it is
necessary to generate the toroidal component of the magnetic field
provided by certain current. It is well known that particles moving
along curved field lines, also drift perpendicularly to the
curvature plane even if the curvature is very small. Such a drift
motion creates curvature current, which leads to the development of
the CDI (see e.g., Ref. \refcite{shmmkh03}).

As we have already mentioned, the innermost region of AGN
magnetospheres is supposed to be rotating, therefore, the
centrifugal force (CF) must be essential for studying the
magnetospheric plasma motion. Magneto-centrifugal effects have been
studied in a series of papers. Blandford and Payne considered the
angular momentum and energy pumping process from accretion disks
(Ref. \refcite{bp82}). The authors emphasized a special role of CF
in dynamical processes governing the acceleration of plasmas. The
role of the centrifugal acceleration in producing the non-thermal
radiation from rotating AGN winds has been studied in Ref.
\refcite{gl97}. This work has been generalized in Ref.
\refcite{osm7} and it has been shown that under certain conditions
electrons might reach very high Lorentz factors $10^{5-8}$. The
similar investigation performed for non-blazar objects was developed
in Ref. \refcite{ra8} and the high efficiency of centrifugal
acceleration was emphasized.

Generally speaking, the centrifugal acceleration may induce plasma
instabilities. When the CF acts on a particle and changes in time,
it plays a role of a parameter, and drives the so-called parametric
instability. Centrifugally excited unstable plasma waves deserve a
special interest in different astrophysical scenarios. In Ref.
\refcite{incr1} the authors considered the corotating Crab
magnetsphere arguing that the centrifugal force inevitably causes
the separation of charges, that in turn leads to the creation of the
Langmuir waves. We have shown that due to the centrifugal effects
the instability is very efficient. This method was applied to AGN
jets in Ref. \refcite{incr3} and the stability problem of the
rotation-induced electrostatic waves has been studied. The CF is
significant also for inducing the CDI. Even if the field lines have
a very small curvature, it might cause a drifting process of plasma,
creating current, which will inevitably produce the toroidal
component of the magnetic field. However, a value of the mentioned
component is not strong enough to change a field line's
configuration significantly. In this context corotation plays an
important role. In particular, as is shown in our model by means of
the centrifugal acceleration, the rotational energy can be pumped
directly into the toroidal component, amplifying it efficiently.
This is the meaning of the curvature drift instability. For
examining the role of the corotation in the curvature drift
instability for pulsar magnetospheres, in Ref. \refcite{mnras} we
studied the two-component relativistic plasma, estimating the
increment of the instability. We have found that the growth rate was
bigger than pulsar's spin-down rate by many orders of magnitude,
indicating high efficiency of the CDI. This instability is very
important since, due to the mentioned drift, current is produced,
which, via the parametric mechanism leads to the creation of the
amplifying toroidal magnetic field, reconstructing the
magnetosphere. By the influence of the toroidal magnetic field, the
initial field lines are twisting until they transform into the shape
of the Archimedes spiral, when the motion of the particles is
described by the so-called force-free regime (see Ref.
\refcite{ff}). The similar problem, thus excitation of the CDI in
AGN magnetospheres was studied in Ref. \refcite{agnff} and it has
been shown that the instability is very efficient and might strongly
influence processes in AGN plasmas. The major restriction of this
work was that we have studied particles with exactly equal initial
phases, and it is clear that most of the electrons have different
values of phases. In the present paper is we address two goals: (a)
we examine the contribution of electrons with different phases in
generation of toroidal magnetic field and (b) study the influence of
the induced magnetic field on dynamics of particles.


The paper is arranged as follows. In Sect. 2, we introduce the
curvature drift waves and derive the dispersion relation and a
corresponding expression of the transition timescale. In Sect. 3,
the results for typical parameters of AGN are presented and, in
Sect. 4, we summarize our results.

\section{Excitation of curvature drift waves} \label{sec:consid}
%
%
%
In AGN magnetospheres the relativistic electrons have energies in a
broad interval. We consider plasma that consists of relativistic
electrons with the following Lorentz factors $\gamma\sim 10^{5-8}$
Ref. \refcite{osm7,ra8}. It is assumed that the field lines
initially are almost rectilinear and corotate. As we have already
mentioned in the introduction, current providing the twisting of
field lines is created due to the curvature drift, characterized by
the following (curvature drift) velocity
\begin{equation}
\label{drift} u= \frac{\gamma_{0} v_{_\parallel}^2}{\omega_B R_B},
\end{equation}
where $\omega_{B} = eB_0/mc$, $e$ and $m$ are particle's charge and
the rest mass respectively, $B_0$ is the unperturbed magnetic
induction, $c$ is the speed of light, $R_B$ is the curvature radius
of magnetic field lines, $\gamma_{0}$ is the initial Lorentz factor
of particles and $v_{_\parallel}$ is the longitudinal velocity. As
it is clear from Eq. (\ref{drift}), the drift velocity is
proportional to $\gamma_0m$, therefore, the corresponding value of
the bulk flow (protons) (having the Lorentz factor of the order of
$10$), will be by many orders of magnitude less than that of the
relativistic electrons (with $\gamma_0\sim 10^{5-8}$). This in turn
means that the contribution of protons is negligible and we consider
one-component plasma composed of relativistic electrons.

In the zeroth approximation, particles move along the magnetic field
lines and our aim is to consider the twisting process of these field
lines and study a subsequent saturation mechanism.

The CDI is centrifugally driven and therefore, for studying the
development of the twisting process of magnetic field lines we
consider the Euler equation, which governs the dynamics of
corotating plasma particles (see Ref. \refcite{incr1})
\begin{equation}
\label{eul} \frac{\partial{\bf p}}{\partial t}+({\bf
v\cdot\nabla)p}= -c^2\gamma\xi{\bf\nabla}\xi+\frac{e}{m}\left({\bf
E}+ \frac{1}{c}\bf v\times\bf B\right),
\end{equation}
where $$ \label{xi} \xi\equiv \sqrt{1-\Omega^2R^2/c^2}, $$ and ${\bf
p}$ is the momentum, ${\bf v}$ is the velocity, ${\gamma}$ is the
Lorentz factor of the relativistic particles, and ${\bf E}$ and
${\bf B}$ are the electric field and the magnetic induction,
respectively. $\Omega$ is the angular velocity of rotation and $R$
is the coordinate along the magnetic field lines. We express the
equation of motion in the cylindrical coordinates (see Fig.
\ref{fig1a}). The first term on the right-hand side of the Euler
equation $-c^2\gamma\xi{\bf\nabla}\xi$ represents the centrifugal
force per unit mass. As we see, this force becomes asymptotically
infinity on the LCS, therefore, its overall effect is significant in
the mentioned zone. For closing the system, we add to Eq.
(\ref{eul}) the continuity equation:
\begin{equation}
\label{cont} \frac{\partial \rho}{\partial t}+{\bf \nabla}\cdot{\bf
J}=0,
\end{equation}
and the induction equation:
\begin{equation}
\label{ind} {\bf \nabla\times B} = \frac{1}{c}\frac{\partial {\bf
E}}{\partial t}+\frac{4\pi}{c}{\bf J},
\end{equation}
respectively. By $\rho\equiv en$ and ${\bf J}\equiv en{\bf v}$ we
denote the charge density and the current density, respectively and
$n$ represents the electron number density.

If we take into account the frozen-in condition, $\bf E_0+ \bf
v_{0}\times\bf B_0/c=0$, describing the leading state of the system,
then one can show that Eq. (\ref{eul}) reduces to \cite{mr}:

\begin{equation}
\label{eul_0} \frac{dv}{dt}=\frac{\Omega^2R}{1-
\frac{\Omega^2R^2}{c^2}}\left[1-\frac{\Omega^2R^2}{c^2}-
\frac{2v^2}{c^2}\right].
\end{equation}
Machabeli \& Rogava have shown that for the ultra relativistic case
($\gamma>>1$) the radial velocity, $v\equiv dR/dt)$ behaves as
follows (see Ref. \refcite{mr})
\begin{equation}
\label{v} v(t)\equiv v_{_\parallel} \approx c\cos(\Omega t+\varphi),
\end{equation}
where $\varphi$ denotes the initial phase of a particle. As we have
already mentioned in the introduction, one of the goals of the
present work is to take into account relativistic electrons with
different values of $\varphi$ and see the corresponding net effect.


We assume that the magnetic field lines initially have a small
curvature, that drives the particles along the $x$ axis (see Fig.
\ref{fig1a}). The drift causes current, which in turn creates the
toroidal component of the magnetic field.

For studying the development of the CDI following the method
described in Reef. \refcite{mnras,ff,agnff} we linearize the system
of equations in Eqs. (\ref{eul}-\ref{ind}), by expanding all
physical quantities around the leading state
\begin{equation}
\label{expansion} \Psi\approx \Psi^0 + \Psi^1,
\end{equation}
\begin{equation}
\Psi = \{n,{\bf v},{\bf p},{\bf E},{\bf B}\},\end{equation}
where $\Psi^0$ and $\Psi^1$ denote the zeroth order and the first
order quantities respectively. We assume that $n_0=const$,
$v_{0x}=u$, $v_{0\theta}=v_{_\parallel}$, ${\bf p_0}={\bf
v_0}/\gamma_0$, $B_{0\theta}=B_0$, $B_{0r}=B_{0x}=0$, $\bf E_0=-\bf
v_{0}\times\bf B_0/c$. $B_0\equiv\sqrt{2L/(R_c^2c)}$ is the
equipartition magnetic field, $L$ is the luminosity of AGN and
$R_{c}=c/\Omega$ is the light cylinder radius.

By expressing the perturbed quantities as
\begin{equation}
\label{pert} \Psi^1(t,{\bf r})\propto\Psi^1(t) \exp\left[i\left({\bf
kr} \right)\right] \,,
\end{equation}
and taking into account $v^1_r\approx cE^1_x/B_{0}$, Eqs.
(\ref{eul}-\ref{ind}) reduce to
\begin{equation}
\label{eulp} \frac{\partial p^1_{x}}{\partial
t}-i(k_xu+k_{\theta}v_{_\parallel})p^1_{x}=
\frac{e}{mc}v_{_\parallel}B^1_{r},
\end{equation}
\begin{equation}
\label{contp} \frac{\partial n^1}{\partial
t}-i(k_xu+k_{\theta}v_{_\parallel})n^1= ik_xn^0v^1_{x},
\end{equation}
\begin{equation}
\label{indp} -ik_{\theta}cB^1_{r} = 4\pi e(n^0v^1_{x}+n^1u),
\end{equation}
where $k_x$ and $k_{\theta}$ are the wave vector components. All
vectors are given in terms of the coordinates of the field line (see
Fig. \ref{fig1a}).

\begin{figure}
\par\noindent
{\begin{minipage}[t]{1.\linewidth}
\includegraphics[width=\textwidth] {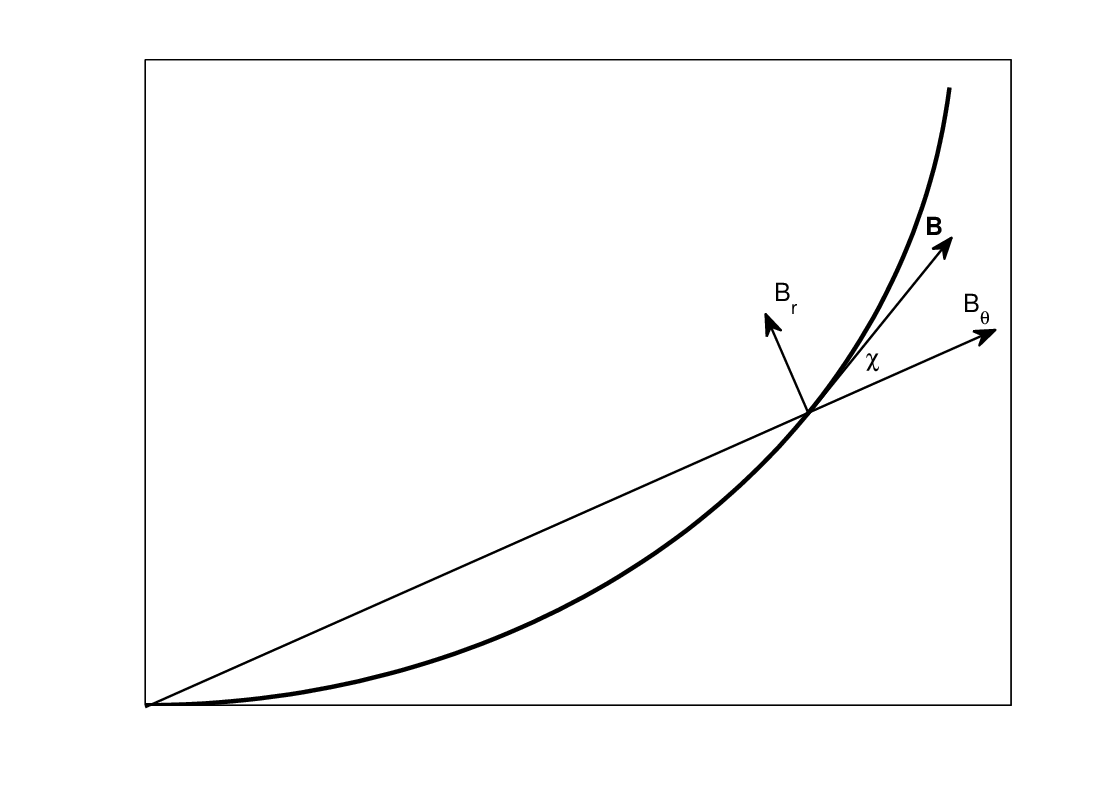}
\end{minipage}
 }
 \caption[ ] {The geometry for deriving Eq. (\ref{time}): the
curved line denotes the twisted magnetic field, ${\bf B}$, generated
due to the raising of magnetic perturbation, ${\bf B}_r$. Note, that
${\bf B}_r$ and ${\bf B}_{\theta}$ are oriented with respect to the
initial `quasi straight' magnetic field line.}\label{fig1b}
 \end{figure}

\begin{figure}
\par\noindent
{\begin{minipage}[t]{1.\linewidth}
\includegraphics[width=\textwidth] {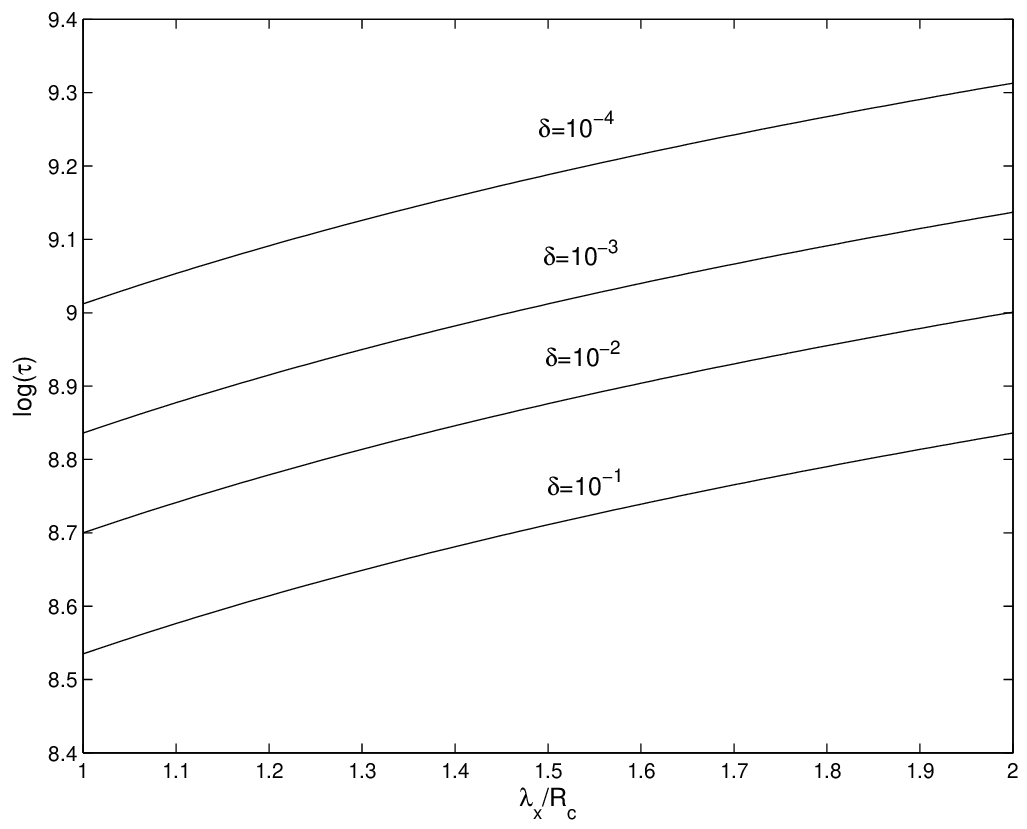}
\end{minipage}
 }
 \caption[ ] {The dependence of logarithm of the transition timescale
on the normalized wavelength. The set of parameters is $\Omega =
3\times 10^{-5}s^{-1}$, $L/L_E = 0.01$, $\delta\in \{10^{-1};
10^{-2}; 10^{-3}; 10^{-4}\}$, $\gamma_{e0} = 10^8$, $R_B\approx
R_{c}$, $n_{e0} = 0.001cm^{-3}$ and $\lambda_{\theta}\equiv
2\pi/k_{\theta} = 100R_{c}$.}\label{fig3}
 \end{figure}

Instead of an ansatz used in Ref. \refcite{agnff}, we apply
\begin{equation}
\label{anzp} v^1_{x}\equiv V_{x}e^{i{\bf kA(t)}},
\end{equation}
\begin{equation}
\label{anzn} n^1\equiv ne^{i{\bf kA}(t)},
\end{equation}

\begin{equation}
\label{Ax} A_{x}(t) = \frac{ut}{2} + \frac{u}{4\Omega}\sin[2(\Omega
t + \varphi)],
\end{equation}
\begin{equation}
\label{Af} A_{\phi}(t) = \frac{c}{\Omega}\sin(\Omega t+\varphi),
\end{equation}
which differs from that of \cite{agnff} by considering the nonzero
initial phases. By combining Eqs. (\ref{anzp}-\ref{Af}) with Eqs.
(\ref{eulp}-\ref{indp}) one gets an expression describing the
development of the toroidal component of the magnetic field
$$ -ik_{\phi}cB^1_{r}(t)
=\frac{\omega^2_{e}}{\gamma_{e0}}{\rm e}^{i{\bf kA}(t)}\int^t{\rm
e}^{-i{\bf kA}(t')}v_{_\parallel}(t')B_{r}(t')dt'+
$$
$$i\frac{\omega^2_{e}}{\gamma_{e0}}k_xu{\rm
e}^{i{\bf kA}(t)}\int^tdt'\int^{t''}{\rm e}^{-i{\bf
kA}(t'')}v_{_\parallel}(t'')B_{r}(t'')dt''$$
\begin{equation}
\label{ind1}
\end{equation}
where $\omega_e = e\sqrt{4\pi n_{e0}/m}$ and $n_{e0}$ are the plasma
frequency and initial electron number density respectively. The
following identity
\begin{equation}
\label{bess} {\rm e}^{\pm ix\sin y}=\sum_s J_s(x){\rm e}^{\pm isy},
\end{equation}
where $J_s(x)$ ($s=0;\pm 1;\pm 2 \ldots$) is the Bessel function of
integer order (see Ref. \refcite{abrsteg})  reduces Eq. (\ref{ind1})
to
$$B_{r}(\omega,\varphi) =
-\frac{\omega_e^2}{2\gamma_{e0}k_{\theta}c}\sum_{\sigma = \pm
1}\sum_{s,q,l,p}\frac{J_s(g)J_q(h)J_l(g)J_p(h)}{\omega +
\frac{k_xu}{2}+\Omega (2s+q) } \times$$ $$\times
B_{r}\left(\omega+\Omega
\left(2[s-l]+q-p+\sigma\right)\right)\times$$
$$\times\left[1-\frac{k_xu}{\omega + \frac{k_xu}{2}+\Omega
(2s+q)}\right]\times{\rm
e}^{i\varphi\left(2[s-l]+q-p+\sigma\right)}+$$
$$+\frac{\omega_e^2k_xu}{4\gamma_{e0}k_{\theta}c}\sum_{\sigma,\mu
= \pm 1}\sum_{s,q,l,p}\frac{J_s(g)J_q(h)J_l(g)J_p(h)}{\left(\omega +
\frac{k_xu}{2}+\Omega (2[s+\mu]+q)\right)^2 } \times$$
$$\times B_{r}\left(\omega+\Omega
\left(2[s-l+\mu]+q-p+\sigma\right)\right)$$
\begin{equation}
\label{disp1}
\;\;\;\;\;\;\;\;\;\;\;\;\;\;\;\;\;\;\;\;\;\;\;\;\;\;\;\;\times {\rm
e}^{i\varphi\left(2[s-l+\mu]+q-p+\sigma\right)},
\end{equation}
where

$$g = \frac{k_xu}{4\Omega}, \;\;\;\;\;\;\;\;\;\;\;\;h =
\frac{k_{\theta}c}{\Omega}.$$

For solving Eq. (\ref{disp1}) we must examine similar expressions,
by rewriting Eq. (\ref{disp1}) for the infinite number of components
$B_r(\omega\pm\Omega,\varphi)$, $B_r(\omega\pm 2\Omega,\varphi)$,...
etc. Therefore, the system becomes composed of the infinite number
of equations, that makes the task unsolvable. On the other hand, by
applying physically reasonable cutoff on the-right hand side of the
equation, the aforementioned problem disappears. In particular, as
we see from the expression the physical system is characterized by
the resonance proper frequency
\begin{equation}
\label{freq}
\;\;\;\;\;\;\;\;\;\;\;\;\;\;\;\;\;\;\;\;\;\;\;\;\;\;\;\;\;\;\;\;\;\;\;\omega_0\approx
-\frac{k_xu}{2},
\end{equation}
$$
\frac{k_xu}{2}<0, \; 2s+q = 0, \; 2[s+\mu]+q = 0.$$

On the other hand, in Ref. \refcite{agnff} it has been shown that
AGN magnetospheric parameters satisfy the condition
$\omega_0/\Omega\ll 1$. This in turn means that all terms with
$\Omega (2s+q)$ and $\Omega (2[s+\mu]+q)$ ($2s+q\neq 0$,
$2[s+\mu]+q\neq 0$) are rapidly oscillating and do not influence the
final result. Therefore, the only contribution comes from $2s+q=
 0$, $2[s+\mu]+q = 0$, which significantly simplifies Eq. (\ref{disp1})

Let us examine an average value of $B_{r}(\omega, \varphi)$ with
respect to $\varphi$, which by taking into account
$$\frac{1}{2\pi}\int{\rm e}^{iN\varphi} d\varphi= \delta_{N,0},
$$
leads to
$$\left(\omega + \frac{k_xu}{2}\right)^2 \approx
$$
\begin{equation}
\label{disp}\approx\sum_{\sigma,\mu = \pm
1}\sum_{s,l}\Xi_{\mu}J_s(g)J_{q'(s,\mu)}(h)J_l(g)J_{p'(l,\sigma)}(h),
\end{equation}
where
\begin{equation}
\label{xi}\Xi_0 = 2\Xi_{\pm 1} =
\frac{\omega^2_{e}k_xu}{2\gamma_{e0}k_{\theta}c},
\end{equation}
$$q' = -2(s+\mu),\;\;\;\;\;\;\; p'=-2l+\sigma.$$
Expressing $\omega\equiv\omega_0+i\Gamma$, one can straightforwardly
derive the increment of the CDI
$$\Gamma\approx$$
\begin{equation} \label{increm} \approx
\left[\sum_{\sigma,\mu = \pm 1}\sum_{s,l}\Xi_{\mu}J_s(g)J_{-2(s +
\mu)}(h)J_l(g)J_{-2l+\sigma}(h)\right]^{\frac{1}{2}}.
\end{equation}


It is worth noting that this instability is unavoidable for plasma
magnetospheric flows, because for the developing of the CDI a) the
initial curvature should be nonzero and b) the magnetic field must
be robust enough to provide the frozen-in condition. (a)-is a
necessary condition for creating the drift waves and (b)-guarantees
the parametric mechanism of rotational energy pumping directly into
the drift modes.

\section{Discussion} \label{sec:results}
%
%
%
Let us consider the Archimedes' spiral $\Phi = aR$, where $\Phi$ and
$R$ are the polar coordinates and $a = {\it const}$. As was shown by
\cite{r03} if the particle slides along the rotating channel that
has the shape of the Archimedes' spiral then, an observer from the
laboratory frame of reference will measure the effective angular
velocity $\Omega_{ef} = \Omega + d\Phi/dt =\Omega + av$, where $v$
is the radial component of velocity of a particle. In case the
motion is force free, thus the particles do not experience any
forces, the trajectory in the laboratory frame of reference will be
a straight line, characterized by the vanishing effective angular
velocity and the following radial velocity, $v = v_c \equiv
-\Omega/a$. This is an interesting property of the Archimedes'
spiral: if one launches a particle along such a rotating channel
(when $a<-\Omega/v_c$), then, if the initial radial velocity exactly
equals $v_c$, the particle will never experience a reaction force,
thus, the dynamics of the particle is always force-free. On the
other hand, a natural question arises: what happens if the initial
velocity differs from $v_c$? As is shown in Ref. \refcite{r03}, the
radial velocity asymptotically behaves according to the following
expression
\begin{equation}
\label{v_r} v_{_R}(R)\rightarrow
-\frac{\Omega}{a}+\frac{E\sqrt{a^2c^2-\Omega^2}}{\Omega c
a^2}\times\frac{1}{R^2},
\end{equation}
where $E$ is the initial energy of the particle per unit of mass.

Typical AGN outflows are highly relativistic, therefore, let us
consider $v_c\approx c$ setting $a = -\Omega /c$. Then, as we see
from Eq. (\ref{v_r}), independently on initial velocity of the the
particle, $v_{_R}$ asymptotically converges to the characteristic
velocity, $c$ ($v_{_R}\rightarrow c$ when $R\rightarrow\infty$).
This implies that dynamics of the particle asymptotically becomes
force-free. The negative sign of $a$ means that the twisting and
rotation have opposite directions and correspondingly electrons
moving along such field lines, will lag behind rotation.

Now we can qualitatively analyze how the configuration of magnetic
field lines changes with time. After perturbing the magnetic field
in the transverse direction, the toroidal component will amplify by
means of the energy pumping directly into the waves, caused by the
parametric nature of the instability. As a result the field lines
will gradually lag behind the rotation. On the other hand, for the
twisted field lines, their dynamical influence on particles will
decrease due to the vanishing reaction force. In the parallel
regime, the efficiency of the CDI will also decrease and when the
field lines get a shape of the Archimedes' spiral the instability
completely vanishes. In particular, for this case
$\Omega_{ef}\rightarrow 0$ and as a result centrifugal effects are
completely damped, the magnetic field lines are saturated,
asymptotically providing a relativistic outflow in the force-free
regime.

Let us assume that when the particles' dynamics becomes force-free,
the critical value of the toroidal magnetic field is $B_r$, then,
referring to Fig. \ref{fig1b}, by taking into account the property
of the Archimedes' spiral, $\tan\chi=1$ (see Ref. \refcite{r03}),
one can show that $\tan\chi = B_r/B_{\theta}$. If we apply the
exponential temporal behaviour of the toroidal component,
$B_r\approx B_r^0\exp (\Gamma\tau)$, then the corresponding
transition timescale gets the form
 \begin{equation}
\label{time} \tau\approx - \frac{1}{\Gamma}\ln
\left(\frac{B^0_r}{B_{\theta}} \right).
\end{equation}

Generally speaking, the role of the CDI is twofold. On the one hand
it guarantees the required twisting of magnetic field lines, and on
the other hand, in terms of its feedback on plasma dynamics it
provides necessary conditions for the saturation process.

We consider the behaviour of the transition timescale versus the
wavelength of the perturbation and the AGN bolometric luminosity
respectively. For this purpose let us examine the following
parameters: $M_{BH}=10^{8}\times M_{\odot}$, $\Omega = 3\times
10^{-5}s^{-1}$ and $L = 10^{44}erg/s$, where $M_{BH}$ and
$M_{\odot}$ are the AGN mass and the solar mass respectively and $L$
is the bolometric luminosity.

In Fig. \ref{fig3} we show the logarithm of the saturation timescale
versus the wavelength normalized to the light cylinder radius for
different values of the initial perturbation
$B^0_r/B_{\theta}\equiv\delta\in \{10^{-1}; 10^{-2}; 10^{-3};
10^{-4}\}$. The set of parameters is $\Omega = 3\times
10^{-5}s^{-1}$, $L/L_E = 0.01$, $\gamma_{e0} = 10^8$, $R_B\approx
R_{c}$, $n_{e0} = 0.001cm^{-3}$ and $\lambda_{\theta}\equiv
2\pi/k_{\theta} = 100R_{c}$, where $L_E = 10^{46}erg/s$ is the
Eddington luminosity for AGN with the given mass. By different
curves we show different cases of initial perturbation. As it is
clear from Fig. \ref{fig3}, the transition timescale varies from
$\sim 10^8s$ ($\lambda_{x}/R_{c} = 1$, $\delta = 10^{-1}$) to $\sim
10^9s$ ($\lambda_{x}/R_{c} = 2$, $\delta = 10^{-4}$). As we see from
the plots, the timescale is a continuously increasing function of
$\lambda_x(\equiv 2\pi/k_x)$, which is natural, because by combining
Eq. (\ref{xi}) with Eq. (\ref{time}) one obtains the following
behaviour $\tau\propto\sqrt{\lambda_x}$.
 \begin{figure}
 \par\noindent
 {\begin{minipage}[t]{1.\linewidth}
 \includegraphics[width=\textwidth] {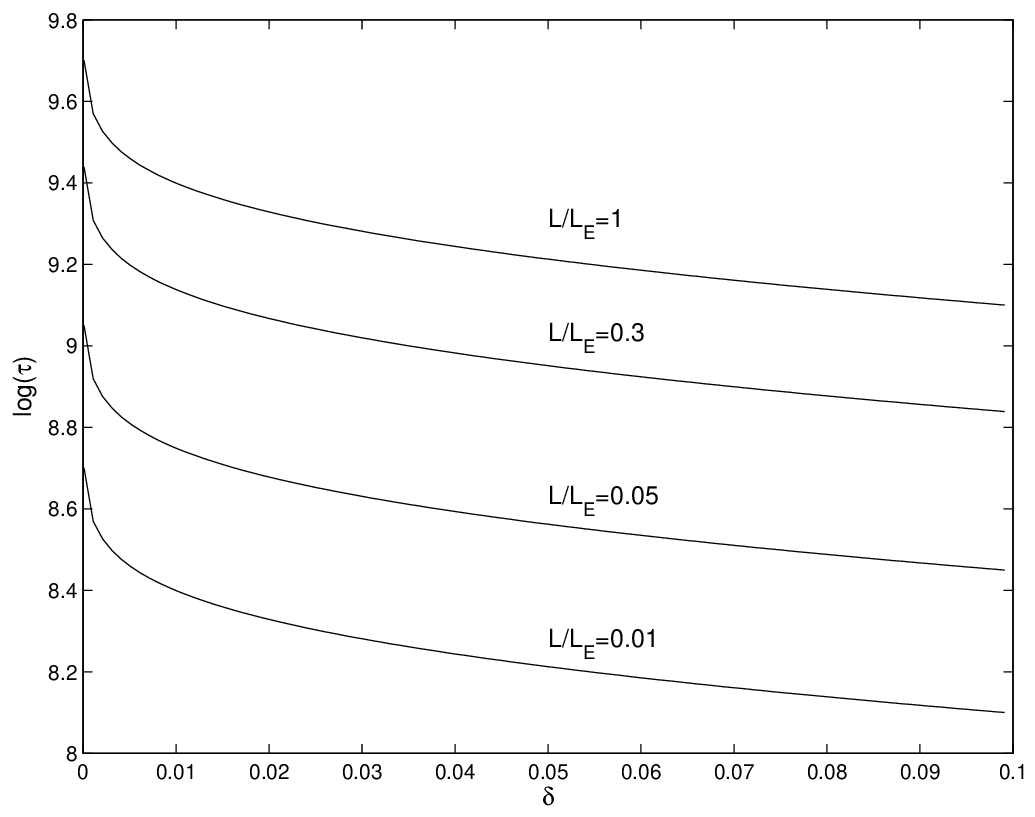}
 \end{minipage}
 }
 \caption[ ] {The dependence of logarithm of the transition timescale
on the dimensionless perturbation. The set of parameters is $\Omega
= 3\times 10^{-5}s^{-1}$, $L/L_E=\{0.01;0.05;0.3;1\}$, $\gamma_{e0}
= 10^8$, $R_B\approx R_{c}$, $n_{e0} = 0.001cm^{-3}$,
$\lambda_{\theta}\equiv 2\pi/k_{\theta} = 100R_{c}$ and $\lambda_x =
R_c$.}\label{fig4}
 \end{figure}

In Fig. \ref{fig4}, we display the behaviour of $log(\tau)$ versus
the initial perturbation for different values of luminosities. The
set of parameters is $\Omega = 3\times 10^{-5}s^{-1}$,
$L/L_E=\{0.01;0.05;0.3;1\}$, $\gamma_{e0} = 10^8$, $R_B\approx
R_{c}$, $n_{e0} = 0.001cm^{-3}$, $\lambda_{\theta}\equiv
2\pi/k_{\theta} = 100R_{c}$ and $\lambda_x = R_c$. The figure shows
the continuously decreasing behaviour of the transition timescale.
This is a natural consequence of the fact that for bigger
perturbations one needs lower time to reach the critical value
$B_r$. On the other hand as we see from the plots, the instability
is less efficient (bigger timescale) for more luminous AGN. Such a
behaviour is clearly seen from an expression of the drift velocity
in Eq. (\ref{drift}). In particular, the drift velocity is
proportional to the inverse value of the cyclotron frequency, which
is bigger for bigger magnetic fields. On the other hand, from an
expression of the equipartition magnetic field we have $B_0\propto
\sqrt{L}$, which means that the curvature drift velocity is
proportional to $1/\sqrt{L}$. Since the curvature drift waves are
more efficient for bigger curvature drift velocities, by increasing
the bolometric luminosity, the increment of the instability will
inevitably decrease and the corresponding transition timescale will
be larger. For the parameters used for Fig. \ref{fig4} the
transition timescale varies from $\sim 10^{10}s$ ($\delta =
10^{-4}$, $L/L_E=1$) to $\sim 10^8s$ ($\delta = 10^{-1}$,
$L/L_E=0.01$).

Summarizing our results we see that the saturation timescale lies in
the range: $\tau\in\{10^8;10^{10}\}s$. A next step is to specify how
efficient is the twisting of field lines. For this purpose it is
relevant to examine an accretion process on AGN, estimate the
corresponding evolution timescale, and compare it with that of the
saturation.

As is shown in Ref. \refcite{king} the accretion timescale is given
by
$$t_{evol}  =3\times
10^{13}\times\left(\frac{\alpha}{0.3}\right)^{-2/27}
\times\left(\frac{\epsilon}{0.1}\right)^{22/27}\times$$
\begin{equation}
\label{tev}\;\;\;\;\;\;\;\;\;\;\times\left(\frac{L}{0.1L_E}\right)^{-22/27}\times
\left(\frac{M_{BH}}{10^8M_{\odot}}\right)^{-4/27}s,
\end{equation}
where $\alpha$ is the Shakura-Sunyaev viscosity parameter (see Ref.
\refcite{sak}), $\epsilon\equiv L/\dot{M}c^2$ is the accretion
parameter and $\dot{M}$ denotes the accretion mass rate.
\begin{figure}
 \par\noindent
 {\begin{minipage}[t]{1.\linewidth}
 \includegraphics[width=\textwidth] {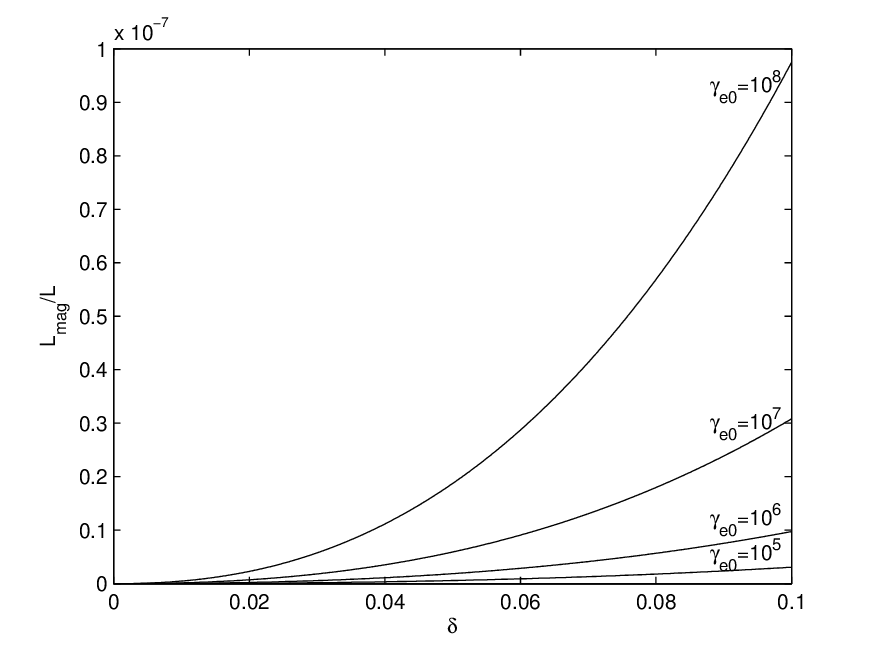}
 \end{minipage}
 }
 \caption[ ] {The dependence of $L_{m}/L_{max}$ versus $\delta$. The set of parameters is
 $\Omega
= 3\times 10^{-5}s^{-1}$, $L = 10^{45}erg/s$, $\gamma_{e0} =
\{10^5;10^6;10^7;10^8\}$, $R_B\approx R_{c}$, $n_{e0} =
0.001cm^{-3}$, $\lambda_{\theta} = 100R_{c}$ and $\lambda_{x} =
R_{c}$.}\label{fig5}
 \end{figure}
For typical values, $\alpha=0.3$, $\epsilon=0.1$, one can see that
the evolution timescale varies from $\sim 10^{12}s$, ($M_9 = 1$ and
$L/L_E = 1$), to $\sim 10^{14}s$ ($M_9 = 0.001$ and $L/L_E = 0.01$),
where $M_{9}\equiv M_{BH}/(10^9\times M_{\odot})$. If one compares
these timescales with that of the transition, one finds that
$t_{evol}$ exceeds $\tau$ by many orders of magnitude, illustrating
the high efficiency of the curvature drift instability.

We have shown that the saturation of the CDI is extremely efficient,
however, it is worth noting that the twisting process of magnetic
field lines requires a certain amount of energy and a natural
question arises: is the energy budget enough to provide the
aforementioned process?.

The maximum value of the energy budget per unit of time, being the
bolometric luminosity itself, $L_{max}=L$ has to be compared with
the "luminosity" corresponding to the twisting of the magnetic field
lines $L_m\equiv \Delta E_m/\Delta t\approx\Delta E_m/\tau$. $\Delta
E_m$ is the variation in the magnetic energy due to the curvature
drift instability.

We consider AGN with $L = 10^{45}erg/s$, then, for the
reconstruction of the magnetosphere one has to satisfy the condition
$L_m<L_{max}$. The magnetic "luminosity" can be estimated
straightforwardly
\begin{equation}
\label{maglum} L_{m} = \frac{B_r^2}{4\pi\tau}R_{c}^3\kappa,
\end{equation}
where $B_r$ behaves as
\begin{equation}
\label{br} B_r = B_r^0e^{t/\tau},
\end{equation}
$B_r^0$ is the initial perturbation of the toroidal component and
$\Delta V\approx R_{c}^2\Delta R=R_{c}^3\kappa$ ($\kappa\equiv
\Delta R/R_{c}<<1$) is the volume, where the twisting process takes
place.

Let us consider the following set of parameters $\gamma_{e0} =
\{10^5;10^6;10^7;10^8\}$, $R_B\approx R_{c}$, $n_{e0} =
0.001cm^{-3}$, $\lambda_{\phi} = 100R_{c}$, $\lambda_{x} = R_{c}$
and $L = 10^{45}erg/s$. In Fig. \ref{fig5} we show the dependence
$L_{m}/L_{max}$ as a function of the initial perturbation for the
moment of the transition ($t\approx\tau$). Different curves
correspond to different Lorentz factors. As it is clear from Fig.
\ref{fig5}, the total luminosity budget, exceeds, by many orders of
magnitude, the magnetic luminosity, indicating that this process is
feasible.

\section{Summary} \label{sec:summary}
%
%
%

We summarize the principal steps and conclusions of our study to be:

\begin{enumerate}

      \item We have studied the curvature drift
      instability and its influence on the dynamics of relativistic
      electrons in AGN magnetospheres.

      \item By linearizing the Euler, continuity and induction equations we have
      derived the dispersion
      relation of the parametrically excited curvature drift
      instability and obtained an expression of the
      growth rate for the light cylinder area. The study was
      performed by generalizing the previous work and taking into
      account different initial phases of relativistic particles.

      \item As a next step we have estimated the transition timescale of
quasi-linear configuration of magnetic field lines into the
Archimedes' spiral. Such a shape of field lines guarantees the
force-free  dynamics of particles and necessarily provides the
saturation of the instability.

      \item We have shown that the transition timescale was
      lower by many orders of magnitude than the accretion evolution
      timescale, illustrating extremely high efficiency of the CDI.

      \item Analyzing the energy budget, it has been shown that the reconstruction
      of the magnetic field lines (which in turn provides the force-free regime of electrons) requires only a tiny fraction of the
      total energy budget, indicating that the CDI is a working
      process and the corresponding transition to force-free dynamics is the realistic process.

\end{enumerate}

In the framework of the present paper two major restrictions have
been considered: (a) a single particle approach and (b) magnetic
field lines located in the equatorial plane. Therefore, it is clear
that a corresponding generalization of the work is needed and we
will investigate it in future studies.

\section*{Acknowledgments}

Z.O. thanks professor G. Machabeli for valuable discussions.

\end{document}